\documentclass[12pt]{article}%

\usepackage{amsfonts}
\usepackage{amssymb}
\usepackage{graphicx}
\usepackage{color}
\usepackage{amsmath}%
\newtheorem{theorem}{Theorem}

\newcommand{\nc}{\newcommand}
\nc{\norm}{\mathcal{N}}
\nc{\gig}{\mathrm{GIG}}
\nc{\ig}{\mathrm{IN}}
\nc{\E}{\mathbb{E}}
\nc{\btheta}{{\bf \theta}}

\nc{\by}{{\bf y}}
\nc{\dd}[2]{\frac{\partial #1}{\partial #2}}
\nc{\lhat}[1][i]{\widehat\lambda_{#1}^{-1(g)}}
\nc{\what}[1][j]{\widehat\omega_{#1}^{-1(g)}}
\nc{\bone}{{\bf 1}}
\nc{\Li}{\widehat\Lambda^{-1(g)}}
\nc{\Oi}{\widehat\Omega^{-1(g)}}
\nc{\diag}[1]{\text{diag}\left(#1\right)}
\nc{\Siginv}{\Sigma^{-1}}
\nc{\Ominv}{\Omega^{-1}}

\def \mb {\mathbf}
\nc{\bx}{\mathbf{x}}
\nc{\D}{\stackrel{\mathrm{D}}{\Rightarrow}}

\begin{document}

\title{\vspace{-0.9cm} Split Sampling:\\
Expectations, Normalisation and Rare Events}

\author{John R. Birge, Changgee Chang, and Nicholas G. Polson\\Booth School of Business\footnote{
        E-mail: john.birge@chicagobooth.edu, changgee@uchicago.edu,
ngp@chicagobooth.edu. We would like to thank the participants at the conference in honor of Pietrio Muliere
at Bocconi, September 13-15, 2012 for their helpful comments. The authors' work was supported by the University of Chicago Booth School of Business.}
        }

\date{First Draft: August 2012\\
This Draft: October 2013}

\maketitle
\begin{abstract}
\noindent In this paper we develop a methodology that we call split sampling methods to estimate high dimensional expectations and rare event probabilities.
Split sampling uses an auxiliary variable MCMC simulation and expresses the expectation of interest as
an integrated set of rare event probabilities. We derive our estimator from a Rao-Blackwellised
estimate of a marginal auxiliary variable distribution.
We illustrate our method with two applications. First, we compute a shortest network path rare event probability and compare our method to estimation to
a cross entropy approach. Then, we compute a normalisation constant of a high dimensional mixture of Gaussians and compare our estimate to one based on
nested sampling.
We discuss the relationship between our method and other alternatives such as
the product of conditional probability estimator and importance sampling.
The methods developed here are available in the {\tt R} package: {\tt SplitSampling}.
\end{abstract}

\vspace{0.1in} {\bf Keywords:} Rare Events, Cross Entropy, Product Estimator,
Slice Sampling, MCMC, Importance Sampling, Serial Tempering, Annealing, Adaptive MCMC,
Wang-Landau, Nested Sampling, Bridge and Path Sampling.

\newpage

\section{Introduction}

In this paper we develop a methodology we refer to as split sampling methods
to provide more precise estimates for high dimensional expectations and rare event probabilities.
We show that more precise estimators can be achieved by splitting the expectation of interest into a number of easier-to-estimate
normalisation constants  and then integrating those estimates to produce an estimate of the full expectation. To do this, we employ
an auxiliary variable MCMC approach with
a family of splitting functions and a weighting function on the conditional distribution of the auxiliary random variable.
We allow for an adaptive MCMC approach to specify our weighting function. We relate our method to the product estimator
(Diaconis and Holmes, 1994, Fishman, 1994) which splits the rare event probability into a set of relatively
larger conditional probabilities which are easier to estimate and to nested sampling (Skilling, 2006) for the estimation of expectations.
Other variance reduction techniques, such as control variates will provide further efficiency gains (see Dellaportas and Kontoyiannis, 2012,
Mira et al, 2012).

There are two related approaches in the literature. One approach is
for estimating expectations is  nested sampling (Skilling, 2006)  which sequentially estimates the
quantiles of the likelihood function under the prior. Other normalisation methods
include bridge and path sampling (Meng and Wong, 1996, Gelman and Meng, 1998), generalized versions of the Wang-Landau algorithm
(Wang and Landau, 2001),
and the TPA algorithm of Huber and Schott (2010). Serial tempering (Geyer, 2010)
and linked importance sampling (Neal, 2005) provide ratios of normalisation constants for a discrete set of unnormalised densities.
The second approach is cross entropy (Rubinstein and Glynn, 2009, Asmussen et al, 2012) which
sequentially constructs an optimal variance-reducing importance function for calculating rare event probabilities. We show that our adaptively chosen weighted
MCMC algorithm can provide efficiency gains over cross entropy methods.

The problem of interest is to calculate an expectation of interest, $ Z= \mathbb{E}_\pi ( L(\bx) ) $, where $L$ is a likelihood and $\pi$ is a prior measure.
For rare event probabilities,
$  Z(m)= \mathbb{E}_\pi ( L_m(\bx) ) $,  and the splitting functions $L_m(\bx)$
are specified by level sets in the likelihood, namely $L_m(\bx) = \mathbb{I} ( L(\bx ) > m )$.
This occurs naturally in the rare event simulation literature and in nested sampling. To develop an efficient estimator, we define a
joint split sampling split distribution, $\pi_{SS}(\bx,m)$, on $\bx$ and an auxiliary variable $m$ that
tilts the conditional distribution with
a weighting function, $\omega(m)$. We provide a default setting for the weight function to match the sampling properties of the product estimator and of nested sampling.
We also allow for the possibility of adaptively learning a weight function from the MCMC output.
As with other ergodic Monte Carlo methods, we assume that the researcher can construct a fast MCMC algorithm for sampling our joint distribution.

The rest of the paper is outlined as follows.
Section 2 details our split sampling methodology.
A key identity ``splits'' the expectation of interest into an integrated set of rare event normalisation constants. MCMC then provides an estimator
of the marginal distribution of the auxiliary variable, which in turn provides our overall estimator.
We provide a number of guidelines for specifying our weight function in both discrete and continuous settings.
Section 3 describes the relationship with the product estimator and nested sampling methods.
In both cases, we provide a theorem that provides a default choice of weight function to match the sampling behavior of the product estimator and
nested sampling. Section 4 applies our methodology to a shortest path rare event probability and to the calculation of
a normalisation constant for a spike-and-slab mixture of Gaussians. We illustrate the efficiency
gains of split sampling over crude Monte Carlo, the product estimator
and the cross entropy method.
Finally, Section 5 concludes with directions for future research.

\section{Split Sampling}

We now introduce the notation to characterize the estimation problems and to develop our method.
The central problem is to calculate an expectation of a non-negative functional of interest, which we denote as $L(\bx)$,
under a $k$-dimensional probability distribution $\pi(\bx)$. We write this expectation as:
$$
Z  = \mathbb{E}_{\pi} \left ( L(\bx) \right ) = \int_{\mathcal{X}} L( \bx) \pi(\bx) d \bx \;.
$$
The corresponding rare event probability is given by
$$
Z(m)  = \mathbb{E}_{\pi} \left ( L_m(\bx) \right ) = \mathbb{P} \left ( L( \bx) > m \right )
$$
We interpret $ \pi( \bx)$ as a ``prior'' distribution, $L(\bx)$ as a likelihood, and $m$
as an auxiliary variable. Given a family of splitting functions, $L_m(\bx)$,
their normalisation constants are, $ Z(m) =  \int_{\mathcal{X}} L_m( \bx) \pi(\bx) d \bx $.
For rare events, $ L_m(\bx) =\mathbb{I} \left ( L(\bx ) > m \right ) $ where $m$ is large.
Here we assume that $L(\bx)$ is continuous with respect to the prior $\pi$ and $Z(0) = \mathbb{P} (L(\bx)>0) = 1$.

Split sampling works as follows.
We define the set of ``tilted'' distributions and corresponding normalisation constants by
$$
\pi( \bx|m) = \frac{L_m(\bx) \pi(\bx)}{Z(m)} \; \; {\rm where} \; \; Z(m) = \int_{\mathcal{X}} L_m(\bx) \pi(\bx) d \bx \; .
$$
For rare events,  $L_m(\bx)=\mathbb{I} \left ( L (\bx ) >m \right ) $ and $\pi(\bx|m) \sim \pi ( \bx | L(\bx) >m )$ corresponds to
conditioning on level sets of $L(\bx)$. The $Z(m)$'s correspond to ``rare'' event probabilities
$$
Z(m) = \int_{\mathcal{X}} L_m( \bx) \pi ( \bx ) d \bx = \int_{ L(\bx) >m } \pi(\bx ) d \bx = \mathbb{P}_{\pi} \left ( L(\bx ) > m \right ) \;.
$$
For the functional $L(\bx)$, the expectation of interest, $Z$, is an integration of
the rare event probabilities. Using Fubini and writing $L(\bx) = \int_0^{L(\bx)} d m $ we have the key identity
$$
Z = \int_{\mathcal{X}} L(\bx) \pi(\bx) d \bx  = \int_{\mathcal{X}} \left \{ \int_{L(\bx)>m} \pi(\bx) dm \right \} d \bx = \int_0^\infty Z(m) d m\; .
$$
We have ``split'' the computation of $Z$ into a set of easier to compute normalisation constants $Z(m)$.
We will simultaneously provide an estimator  $\widehat{Z}(m) $ and $ \widehat{Z}= \int_0^\infty \widehat{Z}(m) d m $.

To do this, we further introduce a weight function on the auxiliary variable, $\omega(m)$, and the cumulative weight
$ \Omega(m) = \int_0^m \omega(s) d s $.
The joint split sampling density, $\pi_{SS} (\bx,m)$, is defined with  $\pi_{SS} \left ( \bx | m \right ) \equiv \pi( \bx |m ) $ as
$$
\pi_{SS} \left ( \bx , m \right ) =\pi_{SS} \left ( \bx | m \right ) \cdot \frac{ \omega(m)Z(m) }{Z_W} \; ,
$$
where $Z_W =  \int_0^\infty \omega(s)Z(s) d s$. The marginals on $m$ and $\bx$ are
\begin{align} \label{marginal_m_x}
\pi_{SS} \left ( m \right ) = \frac{ \omega(m)Z(m) }{ Z_W } \; \; {\rm and} \; \;
\pi_{SS} (\mb{x}) = \frac{\Omega(L(\mb{x})) \pi(\mb{x})}{ Z_W }
\end{align}
where $\Omega(L(\bx)) = \int_0^{L(\bx)} \omega(s) d s  $.

The key feature of our split sampling MCMC draws, $ (\bx,m)^{(i)} \sim \pi_{SS}(\bx ,m) $, are that they provide an
efficient Rao-Blackwellised estimate of the marginal $\widehat{\pi}_{SS}(m) $
without the knowledge of the $Z(m)$'s.
We now show how to estimate $\widehat{Z}(m)$ and $ \widehat{Z} = \int_0^\infty \widehat{Z}(m) dm $.

With $L_m(\bx) = \mathbb{I} \left ( L(\bx) > m \right ) $, the joint splitting density is
$$
\pi(\mb{x},m) = \pi(m) \pi(\mb{x}|M=m) = \frac{\omega(m) Z(m)}{Z_W}
\frac{\mathbb{I} (L(\mb{x})>m ) \pi(\mb{x})}{Z(m)} \qquad 0<m< \infty \; .
$$
The conditional posterior of the auxiliary index $m$ given $\bx$ is
$$
\pi(m|\mb{x}) = \omega(m) \mathbb{I} (m<L(\mb{x}) ) / \Omega (L(\mb{x}))
\; \; {\rm where} \; \; \Omega(m) = \int_0^{m} \omega(s)ds
$$
The density is proportional to the weight function $\omega(m)$ on the interval $[0,L(\bx)]$. Slice sampling corresponds
to $ \omega(m) \equiv 1 $ and uniform sampling on $[0,L(\bx)]$. This would lead to direct draws from the posterior distribution
$ L(\bx)\pi(\bx)/Z$ and the resultant estimator would be the Harmonic mean. Our approach will weight towards regions of smaller $m$ values
to provide an efficient estimator of all the rare event probabilities $Z(m)$ and hence of $Z$.

The marginal density estimator of $m$ is
\begin{align*}
\widehat{\pi}_{SS} (m) & = \frac{1}{N} \sum_{i=1}^N \pi(m|\mb{x}^{(i)}) = \frac{1}{N} \sum_{i=1}^N
\frac{\omega(m) \mathbb{I}{\{L(\mb{x}^{(i)})>m\}}}{\Omega(L(\mb{x}^{(i)}))}, \qquad 0 < m < \infty \\
& = \phi_N( m) \omega( m) \; \; {\rm where} \; \; \phi_N(m) =  \frac{1}{N} \sum_{i=1}^N
\frac{\mathbb{I}{\{ m < L(\mb{x}^{(i)}) \}}}{\Omega(L(\mb{x}^{(i)}))} \; .
\end{align*}
This is a re-weighted version of the initial weights $\omega(m)$. The function $\phi_N(m)$ will be used to adaptively re-balance the initial weights $\omega(m)$
in our adaptive version of the algorithm, see Section 4.

We now derive estimators for $Z(m)$ and $Z$ by exploiting a Rao-Blackwellised estimator for the marginal density, $  \pi_{SS} (m)  $. 
From \eqref{marginal_m_x}, we have 
${\pi}_{SS} (m)  \propto \omega(m) Z(m) $ and so an estimate of $Z(m)$ is given by
$$
\frac{\widehat{Z}(m)}{Z(0)} = \frac{\omega(m)^{-1}\widehat{\pi}_{SS}(m)}{\omega(0)^{-1}\widehat{\pi}_{SS} (0)} \; .
$$
With $Z(0)=1$, this provides a new estimator $\widehat{Z}(m)$, where  $\bx^{(i)} \sim \pi_{SS}(\bx)$, given by
$$
\widehat{Z}(m) = \frac{\phi_N(m)}{\phi_N(0)} =
\frac{\sum_{ i :L(\mb{x}^{(i)})>m} \Omega (L(\mb{x}^{(i)}))^{-1}}{\sum_{i=1}^N  \Omega (L(\mb{x}^{(i)}))^{-1}} \; .
$$
To find $ \widehat{Z} = \int_0^\infty \widehat{Z}(m) d m $, we use the summation-integral counterpart to Fubini and the fact that
$ L(\bx^{(i)} ) = \int_0^{L(\bx^{(i)})} dm $ to yield
$$
\int_0^\infty \sum_{ i :L(\mb{x}^{(i)})>m} \Omega ( L(\mb{x}^{(i)}) )^{-1} d m
 = \sum_{i=1}^N  \Omega \left ( L( \bx^{(i)} ) \right )^{-1} L(\mb{x}^{(i)}) \; .
$$
Therefore, we have our estimator
$$
\widehat{Z} = \sum_{i=1}^N  \frac{ \Omega \left ( L( \bx^{(i)} ) \right )^{-1} }{ \sum_{i=1}^N  \Omega (L(\mb{x}^{(i)}))^{-1}   }  L(\mb{x}^{(i)})    \; .
$$
We now describe our split sampling algorithm.

\vspace{0.2in}
\noindent{\bf Algorithm: Split Sampling}

\begin{itemize}
    \item Draw samples $( \bx ,m)^{(i)} \sim \pi_{SS} \left ( \bx ,m \right )$ by iterating $ \pi_{SS} \left ( \bx |m \right ) $
and $ \pi_{SS} \left ( m|\bx \right )$

  \item Estimate the marginal distribution, $ \widehat{\pi}_{SS} (m)$, via
$$
\widehat{\pi}_{SS} (m) = \frac{1}{N} \sum_{i=1}^N \frac{ \omega(m) L_m(\bx^{(i)}) }{ \int_0^M \omega(m) L_m(\bx^{(i)}) d m } \; .
$$
\item Estimate the individual normalisation constants, $\widehat{Z}(m)$, via
\begin{align} \label{estimator_Z_m}
\widehat{Z}(m) = \frac{\sum_{ i :L(\mb{x}^{(i)})>m} \Omega (L(\mb{x}^{(i)}))^{-1}}{\sum_{i=1}^N  \Omega (L(\mb{x}^{(i)}))^{-1}} \; .
\end{align}
  \item Compute a new estimate, $\widehat{Z}$, via
\begin{align} \label{estimator_Z}
\widehat{Z} = \sum_{i=1}^N  \frac{ \Omega \left ( L( \bx^{(i)} ) \right )^{-1} }{ \sum_{i=1}^N  \Omega (L(\mb{x}^{(i)}))^{-1} } L(\mb{x}^{(i)})    \; .
\end{align}
\end{itemize}

A practical use of the algorithm will involve a discrete grid $0=m_0 < m_1 < \cdots < m_T$. We write the rare event probabilities $Z_t \equiv Z(m_t)$ and the weights $ \omega(m) = \sum_{t=0}^T \omega_t \delta_{ m_t } ( m) $. The
marginal probabilities $\pi_t \equiv \pi(m_t)$ are estimated by Rao-Blackwellization as
$$
\widehat{\pi}_t = \frac{1}{N} \sum_{i=1}^N \pi(m_t|\mb{x}^{(i)}) =
\frac{1}{N} \sum_{i=1}^N \frac{\omega_t \mathbb{I}_{\{L(\mb{x}^{(i)})>m_t\}}}{\sum_{s=0}^T \omega_s \mathbb{I}_{\{L(\mb{x}^{(i)})>m_s\}}},
\qquad  0 \le t \le T.
$$
With $Z_0=1$, the estimator is
$\widehat{Z}_t = \omega_0 \widehat{\pi}_t / \omega_t \widehat{\pi}_0 $ for $ 0 \le t \le T $.
In the next sections we provide a default choice of weights to match the sampling behaviour of the product estimator and nested sampling together
with an adaptive MCMC scheme for estimating the weights.
First, we  turn to convergence issues.

\subsection{Convergence and Monte Carlo standard errors}

Roberts and Rosenthal (1997), Mira and Tierney (2002) and Hobert et al (2002) who provide general conditions for geometric ergodicity of slice sampling.
Geometric ergodicity will imply a Monte Carlo CLT for calculating asymptotic distributions and standard errors.
Our chain is geometrically ergodic if
$ \pi_{SS}$ is bounded, and there exists an $ \alpha > 1 $ such that $G(m)$ is nonincreasing on $(0, \epsilon)$ for some $ \epsilon > 0$ where
$$
G(m) = m^{ \alpha^{-1} +1 } \partial Z(m) / \partial m \; .
$$
Then we can apply a central limit theorem to ergodic averages
of the functional
$$
g(\bx,m) = \Omega ( L( \bx ) )^{-1} \mathbb{I} \left ( L(\bx) > m \right )
$$
which yields the condition
$$
\int_\Theta g(\bx,m)^{2+\epsilon} \pi_{SS} ( \bx ) d \bx =
\int_{ L(\bx) \geq m } \Omega \left ( L(\bx) \right )^{-(1+\epsilon)} \pi ( d \bx ) < \infty
$$
We have a central limit theorem for $\widehat{Z}(m)/Z_W$ at any $u$, where $ 0< \sigma^2_{Z(m)} < \infty $
$$
\frac{\sqrt{N}}{Z_W} \left \{ \widehat{Z}(m) - Z(m) \right \} \D \mathcal{N} \left ( 0 , \sigma^2_{Z(m)} \right )
$$
This argument also works at $m=0$ as long as $ \sigma^2_{Z(0)} < \infty $.

\subsection{Importance Sampling}

The standard Monte Carlo estimate of $ Z = \mathbb{E}_\pi ( L(\bx) ) $ is
$\widehat{Z} = (1/N) \sum_{i=1}^N L( \bx^{(i)} ) $ where $ \bx^{(i)} \sim \pi(\bx) $, with draws possibly obtained via MCMC. This is too inaccurate for rare event probabilities.
Von Neumann's original view of importance sampling was as a variance reduction to improve this estimator.
By viewing the calculation of an expectation as a problem of normalising a posterior distribution, we can write
$$
\pi_L(\bx) = L(\bx) \pi(\bx) / Z \; \; {\rm where} \; \; Z = \int_{\mathcal{X}} L(\bx)\pi(\bx) d \bx \; .
$$
Importance sampling uses a blanket $g(\bx)$ to compute
$$
Z = \int_{\mathcal{X}} L(\bx) \frac{\pi(\bx)}{g(\bx)} g(\bx) d \bx \approx
\widehat{Z}_{IS} = \frac{1}{N} \sum_{i=1}^N  L( \bx^{(i)} ) \frac{\pi ( \bx^{(i)} )}{g( \bx^{(i)})} \; \; {\rm where} \; \;  \bx^{(i)} \sim g(\bx) \; .
$$
Picking $ g(\bx) $ to be the posterior distribution $L(\bx)\pi(\bx)/Z$ leads to the estimator $\widehat{Z}_{IS} = Z$ with zero variance.
While impractical, this suggests finding a class of importance blankets $g(\bx)$ that are adaptive and depend on $L(\bx)$ can
exhibit good Monte Carlo properties.

Split sampling specifies a class of importance sampling blankets, indexed by $\omega(m)$, by
$$
g_\omega (\bx) = \frac{ \left \{ \int_0^{L(\bx)} \omega(s) d s \right \} \pi(\bx) }{ Z_W } = \frac{ \Omega(L(\bx)) \pi(\bx) }{ Z_W }\; .
$$
The estimator $\widehat{Z}(m)$ in \eqref{estimator_Z_m} can be viewed as an importance sampling estimator where we
average $\Omega \left ( L( \bx^{(i)} ) \right )^{-1}$ over the splitting set
$ L( \bx^{(i)} )>m$ with $\bx^{(i)} \sim \pi_{SS}(\bx)$. Similarly we can express $\widehat{Z}$ as an importance sampling estimator as in \eqref{estimator_Z} which uses a proposal distribution proportional to
$ \Omega (L(\bx)) \pi(\bx) $.

\section{Comparison with the Product Estimator and Nested Sampling}

\subsection{Product Estimator}

A standard approach to calculating the rare event probability $Z(m) = \mathbb{P} \left ( L(\bx) > m \right )$ is the product estimator.
We set $m=m_T$ for some $T>0$ and introduce a  discrete grid $m_t$ of $m$-values starting at $m_0=0$. The conditional probability estimator writes
$$
Z(m) = Z_T = \prod_{t=1}^T \mathbb{P} \left ( L(\bx) > m_t |  L(\bx) > m_{t-1} \right )
=  \prod_{t=1}^T\frac{ \mathbb{P} \left ( L(\bx) > m_t  \right )}{ \mathbb{P} \left ( L(\bx) > m_{t-1} \right )}
$$
or equivalently
$Z(m) = Z_T = \prod_{t=1}^T Z_t /Z_{t-1} $.

Variance reduction is achieved by splitting $Z_T$ into pieces $Z_t /Z_{t-1}  $ of larger magnitude which are relatively easier to estimate.
With $ \bx^{(i)}_t \sim \pi_{t-1} (\bx) \equiv \pi(\bx|L(\bx)>m_{t-1})$, we estimate
$$
\frac{Z_t}{Z_{t-1}} =  \int_{\mathcal{X}} L_{m_t}(\bx) \pi_{t-1} (\bx) d \bx
\; \; {\rm with} \; \; \widehat{ \frac{Z_t}{Z_{t-1}} } = \frac{1}{N} \sum_{i=1}^N L_{m_t}( \bx^{(i)}_t  ) \; .
$$
Given $N$ independent samples from the tilted distributions $\pi_{t-1}(\bx)$ for each $t$, we have
$$
\widehat{Z}_T = \prod_{t=1}^T \frac{1}{N} \sum_{i=1}^N  L_{m_t}( \bx^{(i)}_t  ) \;
\; \; {\rm and} \; \;
\frac{\mathrm{Var} \left ( \widehat{Z}_T \right )}{Z_T^2} =  \prod_{t=1}^T
 \left ( \frac{ \sigma_t^2 }{ \mu_t^2 } + 1 \right ) -1
$$
with mean $\mu_t = \mathbb{E} ( \widehat{Z_t / Z_{t-1}} )$ and variance $ \sigma_t^2 = \mathrm{Var} ( \widehat{Z_t / Z_{t-1}} )$.
The product estimator, as well as the
cross-entropy estimator, relies on a set of independent samples drawn in a sequential fashion.
Split sampling, on the other hand, uses a fast MCMC and
ergodic averaging to provide an estimate $\widehat{Z}_T$. The Monte Carlo variation
$ \mathrm{Var} ( \widehat{\pi}_{SS} (m)/\widehat{\pi}_{SS} (0) )  $ can be determined from the output of the chain.
Controlling the Monte Carlo error of this estimator is straightforward due to independent samples with relative mean squared error, see Fishman (1994) and Garvels et al (2002),

\subsubsection{Matching the Product Estimator Sampling Distribution}

We can now compare the product estimator with split sampling.
Suppose $0 < \rho < 1$ and let $m_t$ be the grid points of $L(\bx)$ for $0 \le t \le T$ where $\bx \sim \pi(\bx)$.
By construction, $m_t$ are the $\emph{tail}$ $\rho^t$-quantile of $L(\bx)$, and we have $Z_t = \rho^t$.

There are two versions of the product estimator.
First, the standard product estimator has the long-run sampling distribution
$$
\pi_{PE}(\bx) = \frac{1}{T} \sum_{t=1}^T \pi_{t-1}(\bx) = \frac{1}{T} \sum_{t=1}^T \frac{\pi(\bx)\mathbb{I}(L(\bx) > m_{t-1})}{Z_{t-1}}.
$$
The second product estimator includes samples from the previous level generation that were above the threshold. This has the long-run sampling distribution
\begin{align*}
\pi_{PEI}(\bx) &\propto \pi(\bx) + \sum_{t=2}^T \frac{Z_{t-2}-Z_{t-1}}{Z_{t-2}} \pi_{t-1} (\bx)\\
&= \pi(\bx) + \sum_{t=2}^T \frac{Z_{t-2}-Z_{t-1}}{Z_{t-2} Z_{t-1}} \pi(\bx)\mathbb{I}(L(\bx) > m_{t-1}).
\end{align*}
We call this the product estimator with inclusion.

\begin{theorem}[Product Estimator]
The two product estimators and the split sampling are related as follows.
The standard product estimator corresponds to the (discrete) split sampling with $ \omega_t = 1/Z_t $.
The product estimator with inclusion is equivalent to split sampling with cumulative weights
$$
\Omega_t = \sum_{i=0}^t \omega_i = 1/Z_t.
$$
\end{theorem}

Split sampling with discrete knots $m_t$, weights $\omega_t$ and $\Omega_t = \sum_{i=0}^t \omega_i$ has sampling distribution
\begin{align*}
\pi_{SS}(\bx) &\propto \sum_{t=1}^T \omega_{t-1} \pi(\bx) \mathbb{I}(L(\bx) > m_{t-1})\\
&\propto \sum_{t=1}^{T-1} \Omega_{t-1} \pi(\bx)\mathbb{I}(m_{t-1} < L(\bx) < m_t) + \Omega_{T-1} \pi(\bx)\mathbb{I}(L(\bx) \ge m_{T-1}).
\end{align*}
Therefore, this is equivalent to the product estimator if and only if, for $0 \le t < T$,
$$
\omega_t = \frac{1}{Z_t}, \textrm{ and } \Omega_t = \sum_{i=1}^t \frac{1}{Z_i}.
$$
As $m_t$ are the $\emph{tail}$ $\rho^t$-quantile of $L(\bx)$, we have $\omega_t = \rho^{-t}$ and
$$
\Omega_t = \frac{\rho(\rho^{-t}-1)}{1-\rho}.
$$
For the product estimator with inclusion,
the equivalent split sampling weights are
$$
\Omega_t = 1 + \sum_{i=1}^t \frac{Z_{i-1} - Z_i}{Z_{i-1} Z_i} = \frac{1}{Z_t} = \rho^{-t}.
$$

\subsection{Nested Sampling}

We now provide a comparison with nested sampling. We provide a specification of $\omega(m)$ so that the sampling distribution of matches
that of nested sampling.  We can adaptively determine the values from our MCMC output.
Let $Q_L(q)$ be the $q$-quantile of the likelihood function $L(\bx)$ under the prior $\pi(\bx)$. Then, nested sampling expresses
$$
Z = \int_0^1 Q_L(q) dq = \int_0^\infty Q_L(1-e^{-Y}) e^{-Y} dY.
$$
where $q = 1-e^{-Y}$.
This integral can be approximated by quadrature
$$
Z \approx \sum_{i=1}^\infty (e^{-\frac{i-1}{N}} - e^{-\frac{i}{N}}) Q_L(1-e^{-\frac{i}{N}}).
$$
The larger $N$ is, the more accurate the approximation, and so this suggests the following estimator
$$
\widehat{Z} = \sum_{i=1}^{n-N} (e^{-\frac{i-1}{N}} - e^{-\frac{i}{N}}) L_i + \frac{e^{-\frac{n-N}{N}}}{N} \sum_{i=1}^N L_{n-N+i},
$$
where $L_i$ are the simulated $(1-e^{-\frac{i}{N}})$-quantiles of likelihoods with respect to the prior. Here, $n$ is the total number of samples.
Nested Sampling is a sequential simulation procedure for finding the  $\left( 1-\left(\frac{N}{N+1}\right)^i \right)$-quantiles,
$L_i$, by sampling $ \pi (\bx | L(\bx) > L_{i-1} )$.

Brewer et al (2011) propose a diffuse nested sampling approach to determine the levels $L_t$.
Both nested and diffuse nested sampling are product estimator approaches. The
quantiles $ 0 = L_0 < \ldots < L_t < \ldots  $ are chosen so that each level $L_t$ occupies $\rho = e^{-1} $ times as much prior mass
as the previous level $L_{t-1}$. Diffuse nested sampling achieves this by sequentially sampling from a mixture importance sampler
 $\sum_{j=1}^{t-1} w_j \mathbb{I} \left ( L(\bx) > L_{j-1} \right ) \pi(\bx) $ where the weights are exponential $ w_j \propto e^{ \kappa (j-t)} $ for some $ \kappa$.
MCMC methods are used to traverse this mixture distribution with a random walk step for the index $j$ that steps up or down a level with equal probability.
A new level is added using the $(1-e^{-1})$-quantile of the likelihood draws. Using diffuse nested sampling allows
some chance of the samples' escaping to lowered constrained levels and to explore the space more freely.
One caveat is that a large contribution can come from values of $\bx(z)$ near the origin and we have to find many levels $T$ to obtain an accurate
approximation.


Murray et al (2006) provide single and multiple sample versions of nested sampling
algorithms. If $ L_{\max} = \sup_{ \bx } \; L(\bx ) $ is known, we sample as follows:

\begin{enumerate}
\item Set $ X=1, Z=0,i=1$.
\item Generate $N$ samples $\bx^{(i)}$ from $\pi(\bx) $ and sort $L_i = L( \bx^{(i)} ) $.
\item Repeat while $ L_{max} X > \epsilon Z $:
\begin{enumerate}
\item Set $ Z=Z + L_i X/N $ and $X = ( 1 - 1/N) X $.
\item Generate $  \bx^{(i+N)} \sim \pi(\bx)\mathbb{I} \left ( L(\bx) > L_i \right ) $ and set $ L_{i+N} = L( \bx^{(i+N)} )$.
\item Sort $L_i$'s and set $i=i+1$.
\end{enumerate}
\item Set $Z=Z + (X/N) \sum_{j=1}^N L_{i+j-1} $ and stop.
\end{enumerate}

If $ L_{\max} $ is not known, replace step 3 with:

\begin{enumerate}
\item[3(a).] Repeat while $ L_{i+N-1} X > \epsilon Z $:
\end{enumerate}

\subsubsection{Matching the Nested Sampling Distribution} \label{matching}

We now choose $\omega(m)$ to match the sampling properties of nested sampling. The main difference between split and nested sampling is that
in split sampling we specify a weight function $\omega(m)$ for $0<m<\infty$ and sample from the full mixture distribution, rather than
employing a sequential approach for grid selection which requires a termination rule. Another
difference is that split sampling estimator does not need to know the ordered $L_i$'s.

We can now match the sampling distributions of split and nested sampling (see, Skilling, 2006).
The expected number of samples $L_i$ less than $m$ is
$- N\log Z(m) $.
Assume $N=1$ for a while. Since $Z(L_i)/Z(L_{i-1})$ are independent standard uniforms and
$$
-\log Z(L_k) = -\sum_{i=1}^k \log \frac{Z(L_i)}{Z(L_{i-1})} \qquad \textrm{for } k\ge1 \; ,
$$
the distribution of the number of samples $L_i$ less than $m$ is same as the number of arrivals before $-\log Z(m)$ of a Poisson process with rate $1$.
For general $N$, it only changes the arrival rate of the Poisson process into $N$.

\begin{theorem}[Nested Sampling]
If we pick the weights $\omega(m)$ such that
$$
\Omega(m)=1/Z(m)
$$
Then the sampling distributions of split and nested sampling match.
\end{theorem}

The sampling distribution of the nested sampling for finite $n$  is hard to calculate, but we can observe limiting results.
As $n \rightarrow \infty$, if  $N/n \rightarrow \lambda$, then we have
$$
Z_{NS}(m) = \mathbb{P}^{NS}(L(\bx)>m) = 1 + \lim_{n,N \rightarrow \infty} \frac{N}{n} \log Z(m) = 1 + \lambda \log Z(m).
$$
Split sampling has marginal density of $\bx$ given by
$$
\pi_{SS}(\bx) = \frac{\Omega(L(\bx)) \pi(\bx)}{Z_W} \; \; {\rm with} \; \; \Omega(m) = \int_0^m \omega(s) ds \; .
$$
The tail distribution function $Z_{SS}(m) = \pi_{SS}(L(\bx)>m)$ is then
\begin{align*}
Z_{SS}(m) &= \int_0^\infty \int_{\mathcal{X}}
  \mathbb{I}_{\{L(\bx)>m\}} \frac{\omega(s) Z(s)}{Z_W} \frac{\mathbb{I}_{\{L(\bx)>s\}} \pi(\bx)}{Z(s)} d\bx ds\\
& = \frac{\int_0^\infty \omega(s) Z(m \vee s) ds}{Z_W}
= \frac{Z(m) \Omega(m) + \int_m^\infty \omega(s) Z(s) ds}{Z_W}.
\end{align*}
We now find the importance splitting density
that matches the nested sampling distributional properties in the sense that $Z_{NS}(m) = Z_{SS}(m)$. Since
$$
Z_{NS}^\prime (m) = \frac{\lambda Z^\prime(m)}{Z(m)} \; \; {\rm and} \; \;
Z_{SS}^\prime (m) = \frac{Z^\prime(m) \Omega(m)}{Z_W},
$$
where $ Z_{NS}^\prime (m) = \partial Z_{NS}(m)/\partial m $.
We therefore set $\Omega(m) = 1/Z(m) $.

Since $Z(m)$ is unknown, we are not ready to begin sampling.
As a remedy we propose approximating $Z(m)$.
We estimate $Z_t$'s for certain grid points $\{m_t\}_{0 \le t < T}$
and interpolate $Z(m)$ by, for example, a piecewise exponentially increasing function.
As we assume no information on $L(\bx)$ a priori, we have to start with a single grid point
$m_0=0$ and build more grid points as sampling goes.
When we have collected enough samples higher than the current top grid point,
we add a new grid point and adjust the approximated $Z(m)^{-1}$ function.
For that purpose, we introduce a condition that lets us monitor the
number of visits, $N_{level}$, to the current top level of the likelihood before we construct a new level. Specifically, we run

\begin{enumerate}
\item Set $T=0$, $m_0 = 0$, $\Omega_0 = 1$, and $Z_0=1$.
\item While $T < T_{max}$, set $T=T+1$
\begin{enumerate}
\item Draws $\mb{x}^{(i)} \sim \pi(\bx|L(\bx)>M^{(i-1)})$, and set $L_i = L(\mb{x}^{(i)})$.
\item Obtain $M^{(i)} = \Omega^{-1}(U_i)$ if $U_i>1$ or $M^{(i)} = 0$ otherwise, where $U_i \sim \mathrm{Unif}(0,\Omega(L_i))$.
\item Repeat (a) and (b) until we have $N_{level}$ visits to level $T-1$.
\item Choose the $(1-\rho)$-quantile of likelihoods of level $T-1$ as $m_T$.
\item Set $\widehat{Z}_T = \rho^{-T}$ and $\Omega_T = \widehat{Z}_T^{-1}$.
\end{enumerate}
\end{enumerate}

Under the condition  $\Omega(m) = Z(m)^{-1}$, the chain will visit each level roughly uniformly.
However, it may take a long time to reach the top level, and the uncertainty in $\Omega$ may act like a hurdle for visiting upper levels.
With these concerns, it is desirable to favor upper levels by replacing step (d) with
\begin{enumerate}
\item[]
\begin{enumerate}
\item[(d1)] Set $\Omega_T = e^{\Lambda T} \widehat{Z}_T^{-1}$.
\end{enumerate}
\end{enumerate}
We call $\Lambda$ the boosting factor as $\Lambda$ increases the preference for the  upper levels.
This reduces the search time and ensures the time complexity to be $O(T)$.
To further expedite this procedure, we may put more weight on the top level $T$ by substituting (d) with the step:
\begin{enumerate}
\item[]
\begin{enumerate}
\item[(d2)] Set $\Omega_{T-1} = e^{\Lambda (T-1)} \widehat{Z}_{T-1}^{-1}$ and $\Omega_T = \beta \frac{e^{\Lambda T} -1}{e^\Lambda - 1} \widehat{Z}_T^{-1}$.
\end{enumerate}
\end{enumerate}
For example, if $\beta=1$, the chain spends half of the time on the top level and the other half backtracking the other levels.

Once we identify all levels, our split sampling algorithm runs:

\begin{enumerate}
\item Set $i=0$ and $\nu_t = \nu_{init} \widehat{Z}_t$ for each $t=0,1,\dots,T$.
\item While $i \le n$, set $i=i+1$.
\begin{enumerate}
\item Draw $(\bx,M)^{(i)} \sim \pi_{SS}(\mb{x},m)$ with $\{m_t\}_{0 \le t <T}$ and $\{\Omega_t\}_{0 \le t <T}$, and set $L_i = L(\mb{x}^{(i)})$.
\item For each $t$ with $m_t < L_i$, update $\nu_t = \nu_t + \Omega(L_i)^{-1}$.
\item Update $\widehat{Z}_t = \nu_t/\nu_0$ and set $\Omega_t = \widehat{Z}_t^{-1}$.
\end{enumerate}
\end{enumerate}

From a practical perspective, it is critical to have nonzero initial values on $\nu_t$.
If we start with $\nu_t=0$, the early $\widehat{Z}_t$ and $\Omega_t$ are unstable,
and the whole procedure can become abortive. Since, though not very accurate, $\widehat{Z}_t = \rho^{-t}$ is a reasonable initial estimate,
we use them for the initial values of $\widehat{Z}_t$ and $\Omega_t$.
$\nu_{init}$ reflects the degree of dependence on those initial values.

Another point to make is even if the initial $\Omega_t = \widehat{Z}_t^{-1}$ are not as accurate as
needed to guarantee good mixing of our  MCMC iterations, we
dynamically refine $\Omega_t$ as in step 2(c). The beauty of this algorithm is that this update
makes the chain self-balanced. When $\Omega_t$ is larger than it should be, or $\widehat{Z}_t$ is smaller,
the chain visits level $t$ more often. Thus increasing $\widehat{Z}_t$ and decreases $\Omega_t$, which
helps $\widehat{Z}_t$ converge more quickly to the true value.

At first sight, the time complexity appears to be $O(nT)$ since steps 2(b) and 2(c) cost $O(T)$ operations.
However, if the $m_t$ values are chosen so that $Z_t$ are exponentially decreasing, the work can be done in $O(n\log T)$ time.
The updates needed at steps 2(b) and 2(c) are only for the last several $t$'s since the increment $\Omega(L_i)^{-1}$
becomes negligible very quickly relative to $\nu_t$ as $t$ decreases.

\section{Choice of cumulative weights: $\omega(m)$ and $ \Omega(m)$}

The previous subsection assumed that $\omega(m)$ is fixed. The ``correction factor'' $\widehat{\pi}_{SS} (m)/\widehat{\pi}_{SS} (0)$ in the construction of $\widehat{Z}(m)$
needs to be estimated as accurately as possible.  To do this we will use an adaptive choice of the weight function, $\omega(m)$ and
use convergence results from the adaptive MCMC literature.

A common initialisation is to set $ \omega(m) \equiv 1 , \forall m $ which leads to draws from the posterior.
Then, $ \Omega \left ( L(\bx ) \right ) = \int_0^{L(\bx)} \omega(s) d s = L( \bx) $.
This leads to an estimate of the marginal, $ \mu_N(m) \equiv \widehat{\pi}_{SS}(m) $, given by the measure
$$
 \mu_N (m) = \left \{ \frac{1}{N} \sum_{i=1}^N
\frac{\mathbb{I}{\{ L(\mb{x}^{(i)}) > m \}}}{\Omega(L(\mb{x}^{(i)}))} \right \} \omega(m) \; .
$$
The density estimate will be zero for $m >  \max_i L \left ( \bx^{(i)} \right ) $ by construction.
We also have a set of estimates $ \widehat{Z}(m) = \sum_{ i: L(\bx^{(i)})>m } \Omega(L ( \bx^{(i)} ))^{-1} / \sum_{i=1}^N \Omega(L( \bx^{(i)} ))^{-1} $
where $ \bx^{(i)} \sim \pi_{SS} ( \bx ) $ which can be used to re-balance to weights $ \Omega(m) = Z(m)^{-1}$.

Given an initial run of the algorithm, we can re-proportion the prior weight function to regions we have not visited frequently enough.
To accomplish this, let $ \phi (m) $ be a desired target distribution for $ \pi_{SS}(m) $, for example a uniform measure. Then re-balance
the weights inversely proportional
to the visitation probabilities and set the new weights $\omega^\star (m)$ by
$$
\frac{ \omega^\star (m) }{ \omega(m) } = \frac{ \omega (m )}{ \mu_N(m) } =
 \frac{ \mathbb{I}{\{ L(\mb{x}^{(i)}) > m \}} }{ \Omega(L(\mb{x}^{(i)}))} \; .
$$
This will only adjust our weights in the region where $ m < \max_i L( \bx^{(i)} ) $. As the algorithm proceeds we will
sample regions of higher likelihood values and further adaptive our weight function.

Other choices for weights are available.
For example, in many normalisation problems $Z(m)$ will be exponential in $m$ due to a Laplace approximation argument.
This suggests taking an exponential weighting $\omega(m)=\kappa e^{\kappa m}$ for some $\kappa>0$.  In this case, we have
$$
\Omega(m) = \int_0^{m \wedge M} \omega(s) ds =  e^{\kappa (m \wedge M)}-1 \; .
$$
The marginal distribution is
$$
\pi_{SS} (\mb{x}) = \frac{( e^{ \kappa (L(\mb{x}) \wedge M)} -1) \pi(\mb{x})}{\int_0^\infty \kappa e^{\kappa s} Z(s) ds}.
$$
We can also specify $\omega(m)$ to deal with the possibility that the chain might not have visited all states by setting a threshold
$\omega_{max}$ which corresponds to the maximum allowable increase in the log-prior weights. This leads to a re-balancing rule
$$
\frac{ \omega^\star (m) }{ \omega(m) } = \min \left \{ \frac{ \max_m \mu_N(m) }{\mu_N(m)} , e^{ \omega_{max}} \right \},
$$
where we have also re-normalised the value of the largest state to one.

When $L_{max}$ is available, we set $M=L_{max}$ and $ \omega(m) = 0 $ for $ m>M$.
To initialise $\omega(m)$, we use the harmonic mean
 $ \widehat{Z}^{-1}_{ L_{max} } $ for $\omega(M)$ and an exponential interpolation for $\omega(m)$.
Drawing $ \bx^{(i)} \sim \pi_M(\bx) = L(\bx)\pi(\bx)/Z $, we have
$$
Z^{-1}_M = E_{\pi_M} \left ( L_M^{-1}( \bx) \right ) \; \; {\rm to \; estimate}
\; \; \widehat{\omega}(M) = \frac{1}{N} \sum_{i=1}^N L_M^{-1} (\bx^{(i)})   \; .
$$
The harmonic mean estimator (Raftery et al, 2007) is
known to have poor Monte Carlo error variance properties (Polson, 2006, Wolpert and Schmidler, 2012) although
we are estimating $\omega(m)$ and not its inverse.

We can extend this insight to a fully adaptive update rule for $ \omega_N(m)$, similar to
stochastic approximation schemes. Define a sequence of decreasing positive step sizes
$\gamma_n$ with $ \sum_{n=1}^\infty \gamma_n^{-1} = \infty , \sum_{n=1}^\infty \gamma_n^{-2} < \infty $.
A practical recommendation is $ \gamma_n = C n^{-\alpha} $ where $ \alpha \in [0.6,0.7]$, see e.g. Sato and Ishii (2000).
Another approach is to wait until a ``flat histogram'' (FH) condition holds:
$$
\max_{  m \in \{ m_t \} } \; \vline \; \mu_N ( m ) - \phi (m) \; \vline \; < c  \; .
$$
for a pre-specified tolerance threshold, $c$.
The measure
$\mu_N ( m ) = (1/N) \sum_{i=1}^N \# \left ( m^{(i)} = m \right ) $ tracks our current estimate of the marginal auxiliary variable distribution.
The Rao-Blackwellised estimate $\widehat{\pi}_{SS} ( m ) $ further reduces variance.

The empirical measure can be used to update $\omega_N(m)$ as the chain progresses.  Let $\kappa_N$ denote the
points at which $\gamma_{ \kappa_N } $ will be decreased according to its schedule.
Then an update rule which guarantees convergence is to set
$$
\log \omega_{\kappa_N} ( m ) \leftarrow \log \omega_{\kappa_{N-1}} (m) + \gamma_{\kappa_N} \left ( \mu_{\kappa_N} ( m ) - \phi(m) \right ) \; .
$$
Jacob and Ryder (2012) show that if $\gamma_N$ is only updated on a sequence of values $ \kappa_N $ which correspond to times that
a ``flat-histogram'' criterion is satisfied, then convergence ensues and the FH
criteria is achieved in finite time.
After updating $\omega_{\kappa_N} (m) $, we re-set the counting measure $\mu_{\kappa_N} (m)$ and continue.
Other adaptive MCMC convergence methods are available in Atchade and Liu (2010), Liang et al (2007), and Zhou and Wong (2008).
Bornn et al (2012) provides a  parallelisable algorithm for further efficiency gains.
Peskun (1973) provides theoretical results on optimal MCMC
chains to minimise the variance of MCMC functionals.

One desirable Monte Carlo property for an estimator is a bounded coefficient of variation.
For simple functions, $L(\bx)= \bx $ and $ \max_i \bx_i $, mixture
importance functions achieve such a goal, see Iyengar (1991) and Adler et al (2008). Madras and Piccioni (1999, section 4) hint at the
efficiency properties of dynamically selected mixture importance blankets. Gramacy et al (2010) propose the use of importance tempering.
Johansen et al (2006) use logit annealing implemented via a sequential particle filtering algorithm.

\subsection{Choosing a Discrete Cooling Schedule}\label{cooling}

We suggest a simple, sequential, empirical approach to selecting a ``cooling schedule'' in our approach. Specifically, set
$m_0=0$, then given $m_{t-1}$ we sample  $ \bx^{(i)} \sim \pi_{ m_{t-1} } ( \bx ) \sim \pi \left ( \bx | L(x) > m_{t-1} \right ) $.
We order the realisations of the criteria
function $ L( \bx^{(i)} ) $ and set $ m_t $ equal to the $ ( 1 - \rho )$-quantile of the $L(\bx^{(i)} )$ samples. This provides a sequential
approach to solving
$$
\rho = \mathbb{P} \left ( L(\bx ) > m_t |L(\bx ) > m_{t-1} \right ) =
\mathbb{P} \left ( L(\bx ) > m_t \right )/\mathbb{P} \left ( L(\bx ) > m_{t-1} \right ) = Z_t /Z_{t-1} \; .
$$
A number of authors have proposed ``optimal'' choices of $\rho$, which implicitly defines a cooling schedule, $m_t$, for $0\leq t\leq T$.
L'Ecuyer et al (2006) and Amrein and Kunsch (2011) propose
$\rho = e^{-2} $ and $ 0.2 $, respectively.
Huber and Schott (2010) define a well-balanced schedule as one that satisfies $ e^{-1} < \rho < 2 e^{-1} $.
They show that such a choice leads to fast algorithms.
The difficulty is in finding the right order of magnitude of $M$ and the associated schedule $m_t$ that ensures that each
slice $ Z_t /Z_{t-1}  $ is not exponentially small.
For rare events, we sample until $m_{t+1} > M $ and then set  $ m_{T}=M$. Our initial estimate $\widehat{Z} = \rho^{-T} $ and our weights are
$ \omega (m) = \rho^m $.

In hard cases, such as the multimodal mixture of Gaussians, the normalising constants $Z(m)$ are not exponential in $m$.
In such cases we initialize the weights by a piecewise exponential
obtained by interpolating any point $m \in (m_{t-1},m_t)$ by $\Omega(m) = \Omega_{t-1} \exp( \kappa_t(m-m_{t-1}))$
where $\kappa_t = \log (\Omega_t / \Omega_{t-1}) / ( m_t - m_{t-1} )$. For $m > m_T$, we use $\Omega(m) = \Omega_T$.
The final estimator is given by
$$
\widehat{Z} = \int_0^\infty \widehat{Z}(m)dm = \sum_{t=1}^T \int_{m_{t-1}}^{m_t} \widehat{Z}_{t-1} \exp( -\kappa_t(m-m_{t-1})) dm = \sum_{t=1}^T \frac{(\widehat{Z}_t-\widehat{Z}_{t-1})(m_t-m_{t-1})}{\log \widehat{Z}_t-\log \widehat{Z}_{t-1}}.
$$
Finally, our methodology can be viewed as an adaptive mixture importance sampler. As we rebalanced the weights we are adaptive changing the target distribution of our
MCMC algorithm rather than the traditional adaptive proposal approaches with a fixed target. Other similar approaches include Umbrella sampling (Torrie and Valleau, 1997)
which can be seen as a precursor to many of the current advanced MC strategies
such as the Wang-Landau algorithm and its generalisations for sampling high dimensional multimodal
distributions. These algorithms exploit an auxiliary variable
and by their adaptive nature improve estimates continuously as the simulation advances.
The main difference is how each algorithm traverses low and high energy states. The Wang-Landau
algorithm aims to achieve a uniform distribution on the auxiliary variable, thus spending more time in low energy states than high states states as opposed to
multicanonical sampling (Berg and Neuhaus, 1992), $1/k$-ensemble sampling (Hesselbo and Stinchcombe, 1995)
or simulated tempering (Geyer and Thompson, 1995).

\section{Applications}

\subsection{Rare Event Shortest Path}

Calculating rare event probabilities is a common goal of many problems.
Rubinstein and Kroese (2004)
consider the total length of the shortest path on a weighted graph with random weights $\bx =(x_1 , \ldots , x_5)$. Suppose there are 4 vertices $a$, $b$, $c$, and $d$. The adjacent weight matrix is given by
$$
\left[ \begin{array}{ccccc}
0 & x_1 & x_2 & \infty\\
x_1 & 0 & x_3 & x_4\\
x_2 & x_3 & 0 & x_5\\
\infty & x_4 & x_5 & 0 \end{array}
\right].
$$
Each weight $x_j$ follows an independent exponential distribution with scale parameter $u_j$ with joint distribution given by
$$
\pi (\bx |\mb{u}) = \left (  \prod_{j=1}^5 \frac{1}{u_j} \right ) \exp \left ( - \sum_{j=1}^5 \frac{x_j}{u_j} \right )
\; \; {\rm where} \; \; \mb{u} = (0.25, 0.4, 0.1, 0.3, 0.2 ) \; .
$$
The goal is to estimate the probability of the rare event corresponding to the length of the shortest path from $a$ to $d$
$$
Z(\gamma) = \mathbb{P} (S( \bx ) > \gamma ) \; \; {\rm where} \; \;
S(\bx) = \min ( x_1 +x_4 , x_1+x_3+x_5, x_2+x_3+x_4 , x_2+x_5 ).
$$
We will consider three cases: $\gamma=2,3,$ and $4$ where the true rare event probabilities are
$$
Z(2) = 1.34 \times 10^{-5}, \; \;
Z(3) = 2.06 \times 10^{-8}, \; \; {\rm and} \; \;
Z(4) = 3.10 \times 10^{-11}.
$$
These can be estimated by the split sampler (SS) with $L(\bx) = S(\bx)$ and level breakpoints $\{ 0=m_0, m_1, \dots, m_T=\gamma\}$. $\widehat{Z}_T \equiv \widehat{Z}(m_T)$
is the estimator.

We implement three other competing estimators. First, the crude
Monte Carlo (CMC) estimator simulates $\mb{x}^{(i)} \sim \pi (\mb{x}|\mb{u})$ and estimates the rare event probabilities by
$$
\widehat{Z}(\gamma) = \frac{1}{N} \sum_{i=1}^N \mathbb{I}_{\{ S(\mb{x}^{(i)}) > \gamma \}}.
$$
Second, the conditional probability product (CPP) estimator  $\widehat{Z}(\gamma)$ calculates the $(1-\rho)$-quantile $m_{t+1}$ of $N_0$ samples of $S(\bx^{(t,i)})$
under $\mb{x}^{(t,i)} \sim \pi_t(\mb{x}) \propto \mathbb{I}_{\{ S(\mb{x}) > m_t \}} \pi(\mb{x})$ for all $t=0,\dots,T-1$ with $m_0=0$, $m_{T-1} < \gamma$,
and $m_T \ge \gamma$. This estimator is defined as:
$$
\widehat{Z}(\gamma) = \left( \prod_{t=1}^{T-1} \widehat{ \frac{ Z_t }{ Z_{t-1} } } \right) \widehat{ \frac{ Z( \gamma )}{Z_{T-1} }}
 = \rho^{T-1} \frac{1}{N_0} \sum_{i=1}^{N_0} \mathbb{I}_{\{ S(\mb{x}^{(T-1,i)}) > \gamma \}} \; .
$$
To find $ \bx^{(t,i)} $ we need
to sample $\pi (\mb{x} | S(\bx) > m_t)$. We use Gibbs sampling with complete
conditionals $ \pi ( x_i | \bx_{(-i)} , S(\bx)>m) $ given by truncated exponential distributions.
By the lack of memory property, we have
$$
\pi ( x_1 | \bx_{(-1)} , S(\bx)>m ) = \max ( 0 , m- x_4 , m - x_3 -x_5 ) + x_1^\star \; \; {\rm where} \; \;  x_1^\star \sim Exp(u_1).
$$
The other conditionals $ \pi(x_i | \bx_{(-i)} ,S(\bx)>m) $ follow in a similar manner.

Third, the \emph{cross-entropy} (CE) estimator (de Boer et al, 2005) calculates an ``optimal'' importance blanket,
$ \pi(\bx | \widehat{\mb{v}}_T)$, parameterised by $\widehat{\mb{v}}_T$. Then it draws $N_1$ samples of
$ \mb{x}^{(i)} \sim \pi (\mb{x}|\widehat{\mb{v}}_T)  $ and estimates the shortest path probability
$$
\widehat{Z}(\gamma) = \frac{1}{N_1} \sum_{i=1}^{N_1} \mathbb{I}_{\{ S(\mb{x}^{(i)}) > \gamma \}} w(\mb{x}^{(i)};\mb{u},\widehat{\mb{v}}_T),
\; \; {\rm where} \; \;
w(\mb{x}^{(i)};\mb{u},\widehat{\mb{v}}_T) = \frac{\pi(\mb{x}^{(i)}|\mb{u})}{\pi(\mb{x}^{(i)}|\widehat{\mb{v}}_T)}.
$$
The sequential algorithm for finding $\widehat{\mb{v}}_T$
is similar in spirit to the product estimator approach: set $ \widehat{\mb{v}}_0 = \mb{u} $ and $t=1$. Choose $\rho$; typically $ \rho =0.1$. Then perform

\begin{table}[h!]
\begin{center}
\caption{\label{rare_tbl} Rare event probabilities simulation}
\begin{tabular}{cccccccccccccc}
\hline
 & & \multicolumn{4}{c}{$\gamma=2$} & & \multicolumn{3}{c}{$\gamma=3$} & & \multicolumn{3}{c}{$\gamma=4$}\\
\cline{3-6} \cline{8-10} \cline{12-14}
N & & CMC & CE & CPP & SS & & CE & CPP & SS & & CE & CPP & SS\\
\hline
$10^5$ & & 0.807 & 0.040 & 0.044 & 0.055 & & 0.066 & 0.076 & 0.091 & & 0.113 & 0.098 & 0.133\\
$10^6$ & & 0.275 & 0.011 & 0.015 & 0.015 & & 0.017 & 0.025 & 0.026 & & 0.028 & 0.033 & 0.036\\
$10^7$ & & 0.086 & 0.003 & 0.004 & 0.005 & & 0.005 & 0.007 & 0.007 & & 0.008 & 0.009 & 0.011\\\hline
\end{tabular}
\end{center}
\end{table}

\begin{enumerate}
\item Draw $N$ samples of $ \mb{x}^{(i)} \sim \pi(\mb{x} | \widehat{\mb{v}}_{t-1} ) $. Let $\widehat{\gamma}_t $ be the $(1-\rho)$ quantile of $S(\mb{x}^{(i)})$.

If $\widehat{\gamma}_t > \gamma$, set $\widehat{\gamma}_t = \gamma$.
\item Update $\widehat{\mb{v}}_{t-1} $ via cross-entropy minimisation;
$$
\widehat{\mb{v}}_t = \frac{ \sum_{i=1}^N  \mathbb{I}_{\{S(\mb{x}^{(i)}) > \widehat{\gamma}_t \}} w( \mb{x}^{(i)}; \mb{u} , \widehat{\mb{v}}_{t-1} ) \mb{x}^{(i)} }
{   \sum_{i=1}^N  \mathbb{I}_{\{S(\mb{x}^{(i)}) > \widehat{\gamma}_t \}} w( \mb{x}^{(i)}; \mb{u} , \widehat{\mb{v}}_{t-1} ) }.
$$
\item If $\widehat{\gamma}_t = \gamma$, set $T=t$ and exit. Otherwise, set $t=t+1$ and go to step 1.
\end{enumerate}

\begin{figure}[h!]
\includegraphics[width=\textwidth]{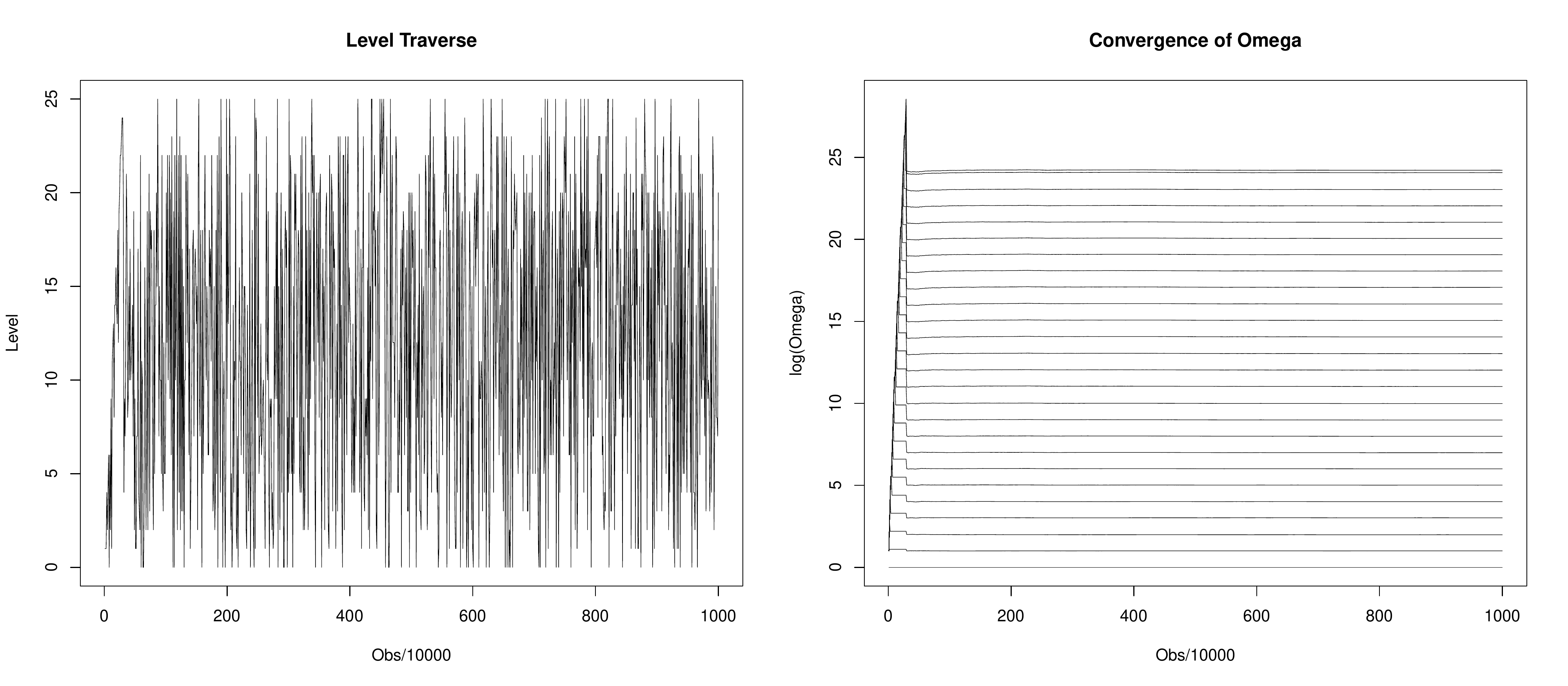}
\caption{The trace plot of visits to likelihood levels and changes of $\Omega_t$ in rare event probability example with $\gamma=4$}
\end{figure}

\begin{figure}[h!]
\includegraphics[width=\textwidth]{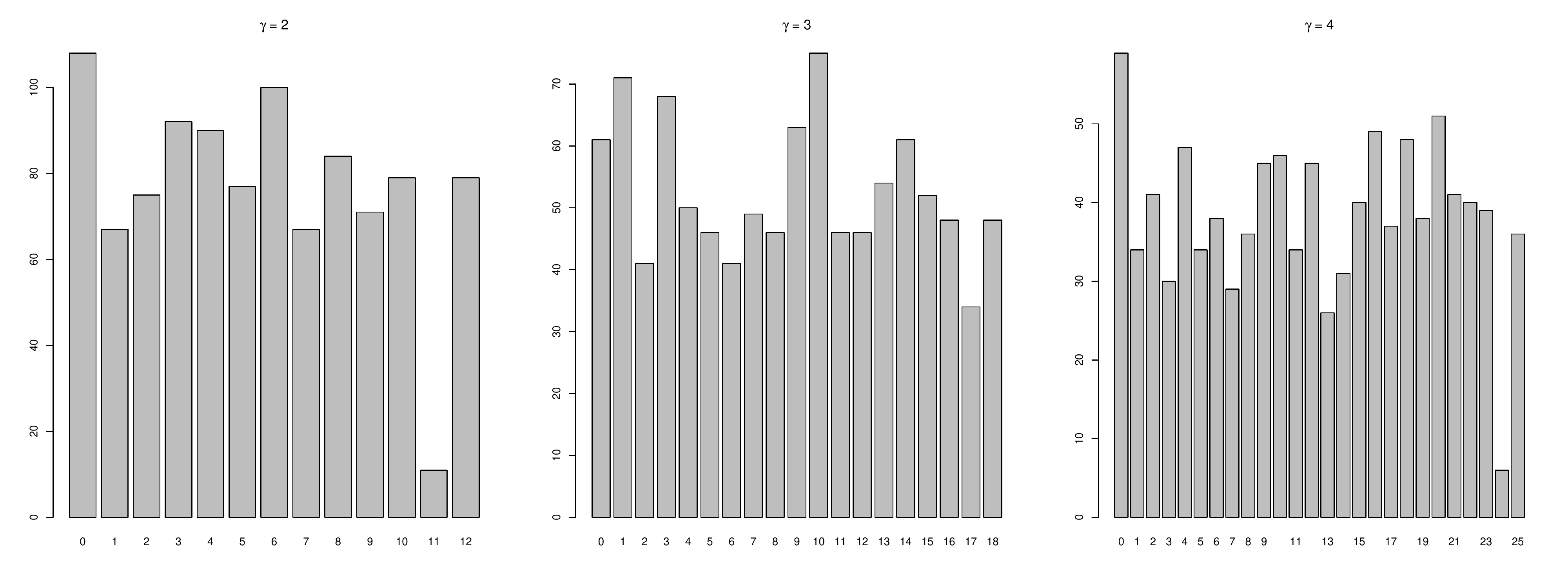}
\caption{Histogram of visits to levels in rare event probability example}
\end{figure}

Table \ref{rare_tbl} provides the simulation results. Each scenario was run $100$ times and
\emph{relative} RMS of each estimator was recorded.
The CMC estimator was only recorded for $\gamma=2$ as the other events are too rare to even have a single count.
The total sample size, $N$, was $10^5$, $10^6$, or $10^7$. The tuning parameters for CE were $\rho=0.1$, $N_0=1,000$ for $\gamma=2$ and $N_0=10,000$ for $\gamma=3,4$, and $N_1 = N - TN_0$.
For CPP, we used $\rho=e^{-1}$ and $N_0 = N/T$ where $T$ is the number of required steps.
For split sampling (SS), we set $\rho = e^{-1}$, $N_{level} = 10,000$ and $\nu_{init} = 10,000$
and $\Lambda = 0.1$.
One can see the cross-entropy method outperforms the others. It is because it finds an efficient importance sampling function and uses independent samples. However, note that the split sampling estimator is almost as efficient as others despite of the fact that it is an MCMC estimator. Further gains from split sampling are expected in higher dimensional problems with multiple modes where finding $ \widehat{\mb{v}}_t $ at each stage in CE
can be cumbersome in general.

\subsection{Normalisation of a Mixture of Gaussians}

As an illustration of the advantages of using split sampling we consider a centered and de-centered mixture of Gaussians.
We follow the nested and diffuse nested sampling literature (Skilling, 2008, Brewer et al, 2011) and suppose that
$ \bx = ( x_1 , \ldots , x_{C} ) $ where $ C=20$. The centered likelihood is given by
the classic Gaussian ``spike-and-slab'' of width $0.01$ and ``plateau'' of width $0.1$, namely
$$
L_C (\bx) = 100 \prod_{i=1}^{20} \frac{1}{\sqrt{2 \pi} u} e^{- \frac{x_i^2}{2 u^2}} +
 \prod_{i=1}^{20} \frac{1}{\sqrt{2 \pi} v} e^{- \frac{x_i^2}{2 v^2}},
$$
The prior $\pi(\bx)$ is uniform on $[-0.5,0.5]^C$.
In the de-centered multimodal mixture we take
$$
L_{DC} (\bx) = 100 \prod_{i=1}^{20} \frac{1}{\sqrt{2 \pi} u} e^{- \frac{(\bx_i - 0.031)^2}{2 u^2}} +
 \prod_{i=1}^{20} \frac{1}{\sqrt{2 \pi} v} e^{- \frac{\bx_i^2}{2 v^2}}.
$$
The goal is to calculate the so-called evidence, $ Z = \int L(\bx ) \pi(\bx) d\bx = 101$.

\begin{figure}[h!]
\includegraphics[width=\textwidth]{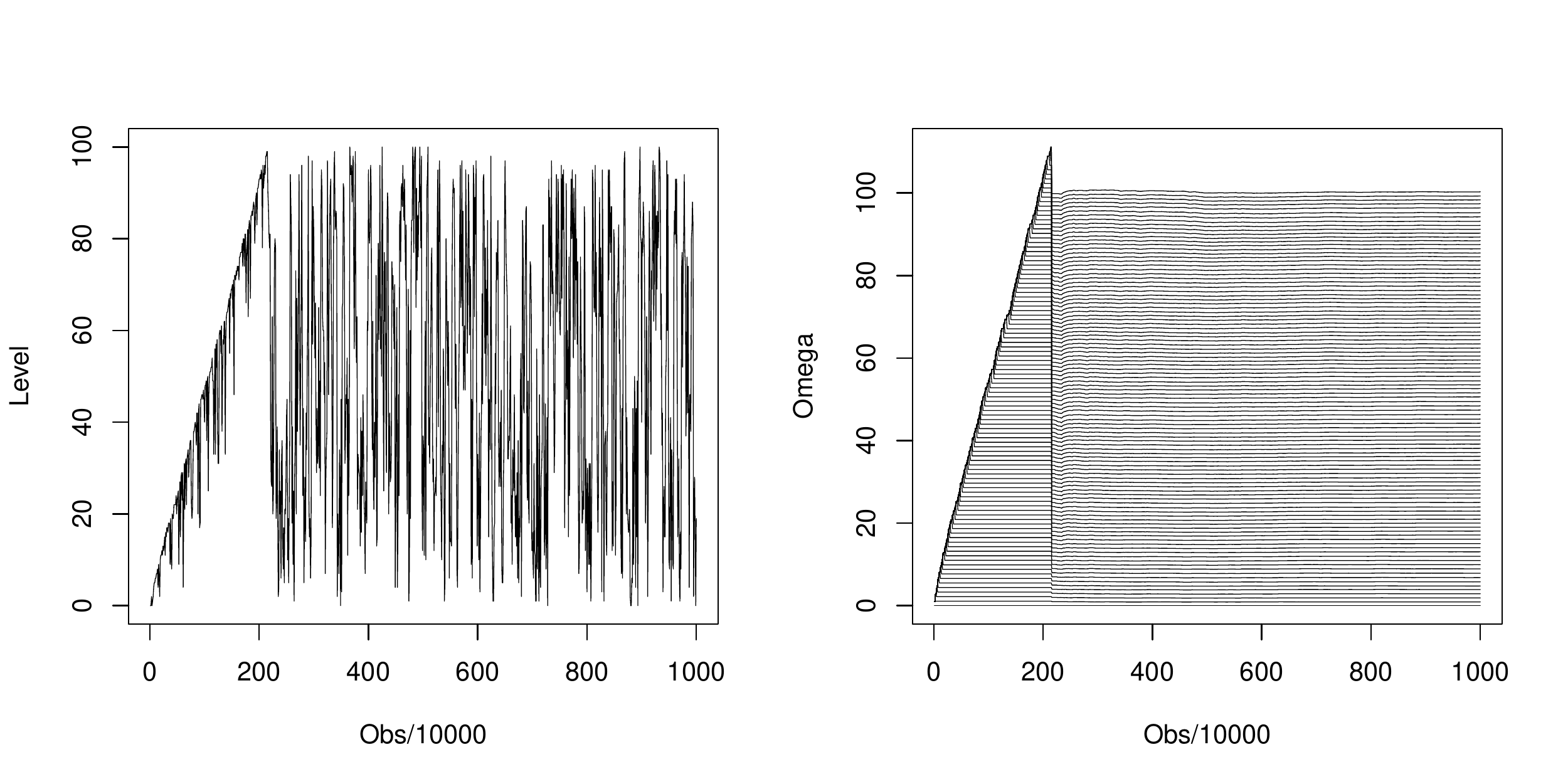}
\caption{The trace plot of visits to likelihood levels and changes of $\Omega_t$ in centered Gaussians case}
\end{figure}

\begin{figure}[h!] 
\includegraphics[width=\textwidth]{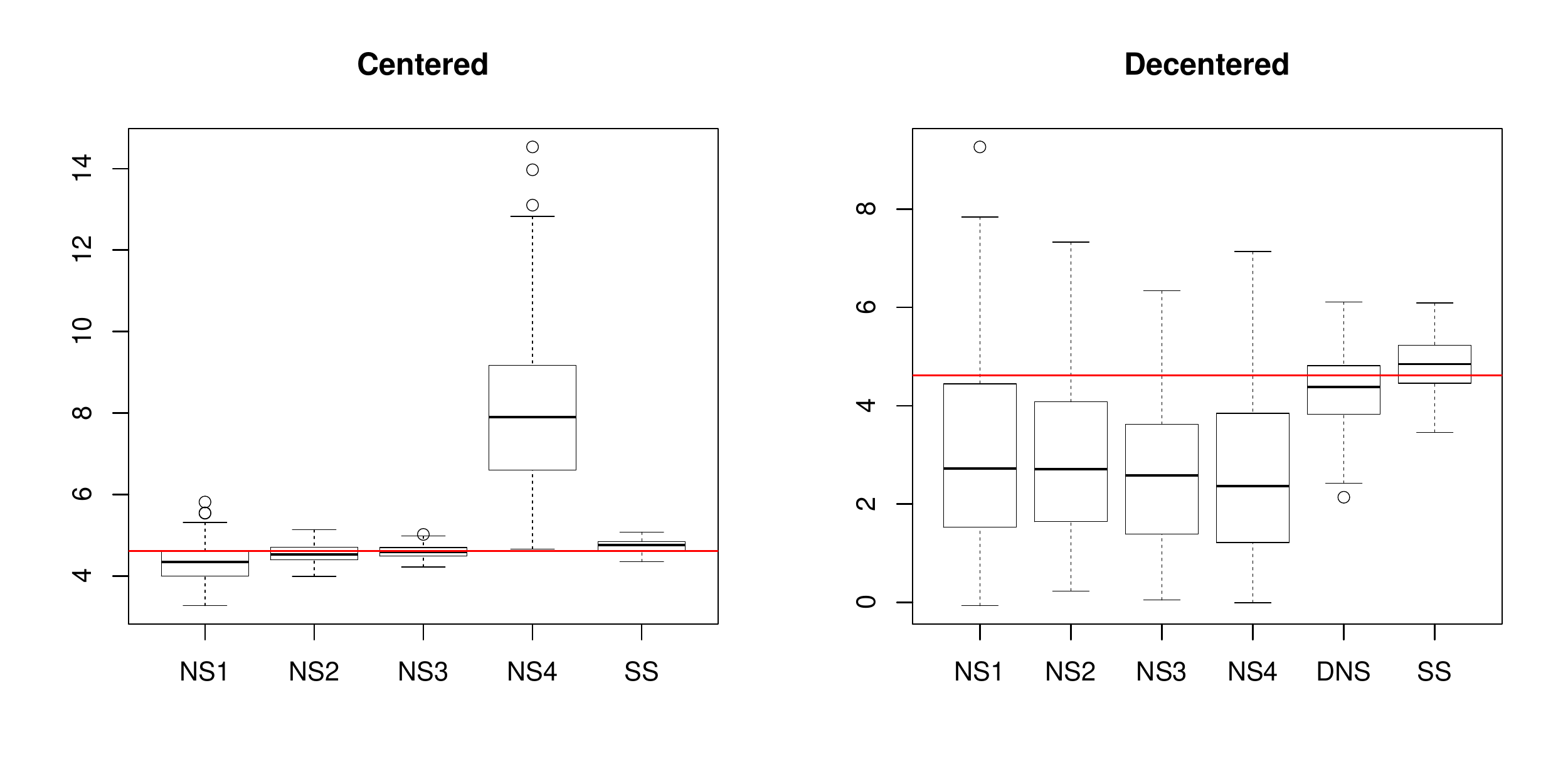}
\caption{\label{boxplot} NS vs DNS vs SS}
\end{figure}

We implemented the split sampling algorithm as described in \ref{matching} and \ref{cooling}.
The nested sampling is implemented via MCMC because drawing from the conditional distribution directly is not possible for this example.
In both methods, we used a random walk MH proposal distribution given by $x_j^\star = x_j + N(0,\sigma^2)$
where the density of the step size $f(\sigma) \propto 1/\sigma$ on $[10^{-4.5},1]$ for random chosen index $j$.

Table \ref{tbl_SS_NS} and the boxplot in Figure \ref{boxplot} compare the performance of the nested sampling and the split sampling methods. For each run, $\log \widehat{Z}$ were recorded and their root mean squares are reported in Table \ref{tbl_SS_NS}. The number of runs for each case is 500. Overall, the nested sampling works slightly better for the centered case.
There is no advantage for split sampling in this case, but it is showing as good performance, too.

\begin{table}[h!]
\begin{center}
\caption{\label{tbl_SS_NS} SS vs NS, $u=0.01$, $v=0.1$, centered at origin, rms of $\log(\widehat{Z})$ with true value $\log(Z) = 4.615$.
The number of MCMC steps is reported per each NS step.}

\begin{tabular}{ccc}
\hline
Algorithm & Parameters & RMS\\
\hline
NS1 & 300 particles, 333 MCMC steps  & $0.557$\\
NS2 & 1000 particles, 100 MCMC steps  & $0.260$\\
NS3 & 3000 particles, 33 MCMC steps  & $0.174$\\
NS4 & 10000 particles, 10 MCMC steps  & $3.647$\\
SS & $\rho=e^{-1}$, $T_{max} = 100$, $\nu=5000$, $\Lambda=10$ & $0.207$\\
\hline
\end{tabular}
\end{center}
\end{table}

\begin{figure}[h!]
\includegraphics[width=\textwidth]{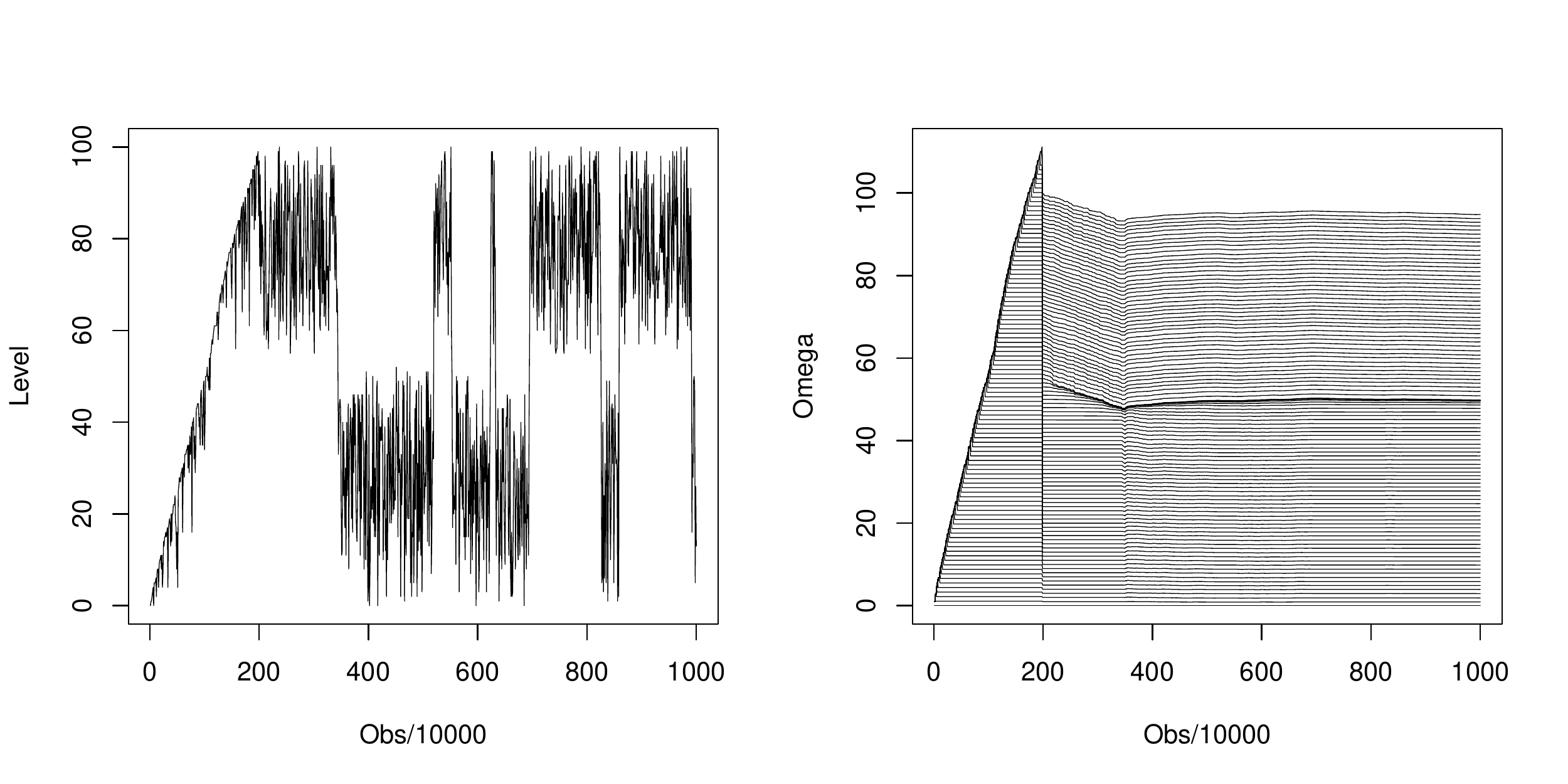}
\caption{\label{decentered} The trace plot of visits to likelihood levels and changes of $\Omega_t$ in de-centered Gaussians case}
\end{figure}

For the de-centered case, the diffuse nested sampling is preferable to the nested sampling,
because when the likelihood is multimodal,
one needs to be able to backtrack the lower levels of likelihood to traverse another mode.
We implemented the diffuse nested sampling for comparison as stated in their paper (Brewer et al, 2011).
As shown in Table \ref{tbl_SS_NS_DNS} and the boxplot in Figure \ref{boxplot}, the split sampling significantly outperforms others.
It is largely due to the fact that the split sampling freely visits the two modes of likelihood function
and fast convergence of $\Omega_t$'s and $\widehat{Z}_t$'s as one can see in the plots in Figure \ref{decentered}.

\begin{table}[h!]
\begin{center}
\caption{\label{tbl_SS_NS_DNS} SS vs (D)NS, $u=0.01$, $v=0.1$, the spike centered at $(0.031, \dots, 0.031)$, rms of $\log(\widehat{Z})$ with $\log(Z) = 4.615$.
The number of MCMC steps is per each NS step. Diffuse nested sampling (DNS, Brewer et al, 2011).}

\begin{tabular}{ccc}
\hline
Algorithm & Parameters & RMS\\
\hline
NS1 & 300 particles, 333 MCMC steps  & $2.467$\\
NS2 & 1000 particles, 100 MCMC steps  & $2.338$\\
NS3 & 3000 particles, 33 MCMC steps  & $2.519$\\
NS4 & 10000 particles, 10 MCMC steps  & $2.620$\\
DNS & Diffuse Nested Sampling & $0.763$\\
SS & $\rho=e^{-1}$, $T_{max} = 100$, $\nu=5000$, $\Lambda=10$ & $0.591$\\
\hline
\end{tabular}
\end{center}
\end{table}

\section{Discussion}

The advantage of the class of split sampling densities
is that the resultant estimator of $Z$  can be implemented via an auxiliary MCMC algorithm from a joint distribution
$ \pi_{SS} ( \bx , m ) $ indexed by a random auxiliary variable $ m$. Moreover, it allows an adaptive choice of
$\omega_N(m)$ to reduce the Monte Carlo error of the resultant estimator.
Convergence results rely on adaptive MCMC literature.
Roberts (2010) observes that MCMC methods are likely to achieve the largest efficiency gains in rare event probabilities
(Glasserman et al, 1999, Glynn et al, 2010)
and in counting problems where the resultant chains can be hard to sample exactly.

Split sampling illustrates the adaptive importance sampling nature of nested sampling and cross-entropy methods.
There is also a clear relationship with slice sampling (Polson, 1996, Neal, 2003) as one can view the sampling of the posterior,
$ \pi_L(\bx)$, as the marginal from the augmented distribution $ \pi ( \bx , m ) = \mathbb{I} \left ( L(\bx) > m \right ) \pi(\bx) / Z $.
The main difference is that split sampling runs a Markov chain that
traverses the whole space defined by $(\bx , m)$ to
find regions where $\omega(m)$ needs to be refined. Both CE and NS methods using a sequential sampling procedure as in the CPP estimator
to split the quantity of interest into estimable pieces.
Further research is required to tailor the specification of the weight function $\omega(m)$ to the problem at hand.

We leave open the question of an optimal choice of $L_m(\bx)$. Here we have focused on $L_m(\bx) = \mathbb{I}(L(\bx)>m)$, however,
using logit-type functions might led to faster converging MCMC algorithms.
The key to the efficiency of split sampling is being able to construct
a rapidly mixing MCMC algorithm to sample the mixture distribution $\pi_{SS}(\bx,m)$.
We aim to report on direct applications in Bayesian inference in future work.
For example, Murray et al (2006) shows that nested sampling performs well for Markov random fields models and split sampling
should have similar properties.

\section{References}
\noindent Adler, R.J., J. Blanchet and J. Liu (2008). Efficient simulation for tail probabilities of Gaussian random fields.
\textit{Proceedings of the Winter Simulation Conference}, 328-336.\medskip

\noindent Amrein, M. and H. K\"unsch (2011). A variant of importance sampling for rare event estimation: fixed number of successes.
\textit{ACM Transactions on Modeling and Computer Simulation}, 21(2), Article 13.\medskip

\noindent Asmussen, S. and P. Dupuis, R. Rubenstein and H. Wang  (2012). Importance Sampling for Rare Events.
\textit{Encyclopedia of Operations Research} (eds Gass, S. and M. Fu), Kluwer.\medskip

\noindent Atchad\'e,Y.F. and J.S. Liu (2010). The Wang-Landau algorithm in general state spaces: applications and convergence analysis.
\textit{Statistica Sinica}, 20, 209-233.\medskip

\noindent Berg, B.A. and T. Neuhaus (1992). Multicanonical ensemble: A new approach to simulate first-order phase transitions.
\textit{Physical Review Letters}, 68, 9-12.\medskip

\noindent Bornn, L., P.E. Jacob, P. Del Moral and A. Doucet (2012). An adaptive Wang-Landau algorithm for automatic density exploration.
\textit{J. Computational and Graphical Statistics}.\medskip

\noindent de Boer, P.T., D.P. Kroese and R.Y. Rubenstein (2005). A Tutorial on the cross-entropy method.
\textit{Annals of Operations Research}, 134, 19-67.\medskip

\noindent Brewer, B.J, L.B. P\'artay and G. Cs\'anyi (2011). Diffusive Nested Sampling.
\textit{Statistics and Computing}, 21(4), 649-656.\medskip

\noindent Dellaportas, P. and I. Kontoyiannis (2012). Control variates for estimation based on reversible MCMC samplers.
\textit{Journal of Royal Statistical Society, B}, 74(1), 133-161.\medskip

\noindent Diaconis, P. and S. Holmes (1994).
Three examples of the MCMC method. \textit{Discrete Probability and Algorithms}, 43-56.\medskip

\noindent Fishman, G. (1994). Markov chain sampling and the Product Estimator.
\textit{Operations Research}, 42(6), 1137-1145.\medskip


\noindent Garvels, M.J.J., J.C.W. van Ommeren and D.P. Kroese (2002). On the importance function in splitting simulation.
\textit{European Transactions on Telecommunications}, 13(4), 363-371.\medskip

\noindent Gelman, A. and X. Meng (1998). Simulating normalizing constants: from importance sampling
to bridge sampling to path sampling. \textit{Statistical Science}, 13, 163-185.\medskip

\noindent Geyer, C.J. and E.A. Thompson (1995). Annealing Markov chain Monte Carlo with applications
to ancestral inference. \textit{Journal of  American Statistical Association}, 90, 909-920.\medskip

\noindent Geyer, C.J (2012). Bayes factors via Serial Tempering.
\textit{Technical Report}, University of Washington.\medskip

\noindent Glasserman, P., P. Heidelberger, P. Shahabuddin and T.Zajic (1999).
Multi-Level Splitting for rare event probabilities. \textit{Operations Research}, 47(4), 585-600.\medskip

\noindent Glynn, P.W., A. Dolgin, R.Y. Rubinstein and R. Vaisman (2010). How to generate uniform samples on discrete sets using
the splitting method. \textit{Prob. Eng. Info. Sci.}, 24(3), 405-422.\medskip

\noindent Gramacy, R., R. Samworth and R. King (2010). Importance Tempering. \textit{Statistics and Computing}, 20(1), 1-7.\medskip

\noindent Hesselbo, B. and R.B. Stinchcombe (1995). Monte Carlo Simulation and global optimisation without parameters.
\textit{Phys. Rev. Lett.}, 74, 2151-2155.\medskip


\noindent Huber, M. and S. Schott (2010). Using TPA for Bayesian Inference.  \textit{Bayesian Statistics, 9}, 257-282.\medskip

\noindent Iyengar, S. (1991). Importance Sampling for Tail Probabilities.
\textit{Technical Report} 440, Stanford University.\medskip

\noindent Jacob, P.E. and R.J. Ryder (2012). The Wang-Landau algorithm reaches the Flat Histogram criterion in finite time.
\textit{Ann. Appl. Probab.}.\medskip

\noindent Johansen, A.M., P. Del Moral and A. Doucet (2006). Sequential Monte Carlo Samplers for Rare Events.
\textit{Proceedings of the 6th International Workshop on Rare Event Simulation}, 256-267.\medskip

\noindent L'Ecuyer, P., V. Demers and B. Tuffin (2006). Splitting for Rare event simulation.
\textit{Proceedings of the 2006 Winter Simulation Conference}, 137-148.\medskip

\noindent Liang, F. (2005). Generalized Wang-Landau algorithm for Monte Carlo computation.
\textit{Journal of American Statistical Association}, 100, 1311-1337.\medskip

\noindent Liang, F., C. Liu and R.J. Carroll (2007). Stochastic approximation in Monte Carlo computation.
\textit{Journal of American Statistical Association}, 102, 305-320.\medskip

\noindent Madras, N. and M. Piccioni (1999). Importance Sampling for Families of Distributions.
\textit{Annals of Applied Probability}, 9(4), 1202-1225.\medskip

\noindent Meng, X-L and W. Wong (1996). Simulating ratios of normalising constants via
a simple identity: a theoretical exposition.
\textit{Statistia Sinica}, 6, 831-860. \medskip

\noindent Mira, A., R. Solgi, and D. Imparato (2012). Zero Variance MCMC for Bayesian Estimators.
\textit{Statistics and Computing}, 23(5), 653-662.\medskip

\noindent Murray, I., D.J.C. MacKay, Z. Ghahramani and J. Skilling (2006). Nested sampling for the Potts models.
\textit{Advances in NIPS}, 947-954.\medskip

\noindent Neal, R.M. (2003). Slice Sampling. \textit{Annals of Statistics}, 31(3), 705-767.\medskip

\noindent Neal, R.M. (2005). Estimating ratios of normalising constants using linked importance sampling.
\textit{Technical Report} No. 0511, University of Toronto.\medskip

\noindent Peskun, P.H. (1973). Optimum Monte Carlo sampling using Markov Chains. \textit{Biometrika}, 60(3), 607-612.\medskip


\noindent Polson, N.G. (1996). Convergence of Markov Chain Monte Carlo Algorithms.
\textit{Bayesian Statistics, 5}, 297-321.\medskip

\noindent Polson, N.G. (2006).  Comment on: ``Estimating the integrated likelihood in posterior simulation
using the harmonic mean equality''. \textit{Bayesian Statistics, 8}, 415-417.\medskip

\noindent Raftery, A.E., M.A. Newton, J.M. Satagopan and P.N. Krivitsky (2007). Estimating the integrated likelihood via posterior simulation
using the Harmonic mean equality. \textit{Bayesian Statistics, 8}, 371-417.\medskip

\noindent Roberts, G.O. (2010). Comment on: ``Using TPA for Bayesian Inference''. \textit{Bayesian Statistics, 9}, 280-282.\medskip

\noindent Rubinstein, R.Y. and P.W. Glynn (2009). How to deal with the curse of dimensionality of likelihood ratios in Monte Carlo simulation.
\textit{Stochastic Models}, 25, 547-568.\medskip

\noindent Rubenstein, R.Y. and D. Kroese (2004). \textit{The Cross Entropy Method}. Springer.\medskip

\noindent Sato, M. and S. Ishii (2000). On-line EM algorithm for the normalized Gaussian network. \textit{Neural Computation}, 12, 407-432.\medskip

\noindent Skilling, J. (2006). Nested Sampling for General Bayesian Computation. \textit{Bayesian Analysis}, 1(4), 833-860.\medskip

\noindent Skilling, J. (2008). Nested Sampling for Bayesian computation. \textit{Bayesian Statistics, 8}, 491-507.\medskip

\noindent \v{S}tefankovi\v{c}, D., S. Vempola and E. Vigoda (2009). Adaptive simulated annealing: a near-optimal connection between sampling and counting.
\textit{Journal of the ACM}, 56(3), Article No. 18.\medskip

\noindent Torrie, G.M. and J.P. Valleau (1977). Nonphysical sampling distributions in
Monte Carlo free-energy estimation: Umbrella Sampling. \textit{Journal of Computational Physics}, 23(2), 187-199.\medskip

\noindent Wang, F. and D.P. Landau (2001). Efficient multiple-range random walk algorithm to calculate the density of states.
\textit{Phys. Rev. Lett}, 86, 2050-2053.\medskip

\noindent Wolpert, R.L. and S.C. Schmidler (2012). $\alpha$-Stable limit laws for Harmonic Mean estimators of marginal likelihoods.
\textit{Statistica Sinica}, 22, 1233-1251.\medskip

\noindent Zhou, Q. and W.H. Wong (2008). Reconstructing the energy landscape of a distribution from Monte Carlo samples.
\textit{Annals of Applied Statistics}, 2(4),1307-1331.\medskip

\end{document}